\documentstyle[prl,aps,epsfig]{revtex}

\newcommand{\bea}{\begin{eqnarray}}
\newcommand{\eea}{\end{eqnarray}}

\def\be{\begin{equation}}
\def\ee{\end{equation}}

\begin{document}  
\draft
\twocolumn[\hsize\textwidth\columnwidth\hsize\csname
@twocolumnfalse\endcsname
\preprint{ }

\title{Unified Explanation of the Solar and Atmospheric neutrino Puzzles 
in a supersymmetric SO(10) model}

\author{B. Brahmachari and R. N. Mohapatra}

\address{{\it Department of Physics, University of Maryland,
College Park, MD 20742.}}

\maketitle

\begin{abstract} 
It was recently suggested that in a class of supersymmetric 
SO(10) models with Higgs multiplets in ${\bf 10}$, and a single ${\bf 126 
+ \overline{126}}$ representations, if the ${\bf \overline{126}}$ 
contributes both to the right handed neutrino masses as well as to the 
charged fermion masses, one can have a complete prediction of the 
neutrino masses and mixings. It turns 
out that if one chooses only one {\bf 10}, there are no regions in the
parameter space where one can have a large $\nu_{\mu}-\nu_{\tau}$ mixing
angle necessary to solve the atmospheric neutrino deficit while at the 
same time solving the solar neutrino puzzle via the $\nu_e 
\leftrightarrow \nu_{\mu}$
oscillation. We show that this problem can be solved in a particular
class of SO(10) models with a pair of {\bf 10} multiplets if we 
include the additional left-handed triplet contribution to the light 
neutrino mass matrix. This model cannot reproduce the mass and mixing 
parameters
required to explain the LSND observations neither does it have
have a neutrino hot dark matter.\\
{\bf UMD-PP-98-49}
\end{abstract}
\vskip1pc]
Strong indications in favor of non-vanishing neutrino masses are 
emerging from several experiments: (i) the deficits of solar neutrino 
flux observed by the four solar neutrino experiments Homestake, 
Kamiokande, SuperKamiokande, SAGE and GALLEX\cite{solar} compared to the 
standard solar model calculations \cite{bahcall} can be understood if 
neutrinos are massive and the electron neutrinos emitted by the sun 
oscillate to another neutrino species. and (ii) the atmospheric 
muon neutrino deficits observed earlier by Kamiokande, IMB and Soudan 
II\cite{soud} experiments and confirmed recently by Super Kamiokande can 
be understood if $\nu_\mu$ oscillates similarly. The LSND\cite{lsnd} 
results have provided the first laboratory indication of 
$\overline{\nu}_\mu \leftrightarrow \overline{\nu}_e$ oscillation and if 
confirmed by KARMEN\cite{karmen}, would seal the case for non-zero 
neutrino masses, in an unequivocal manner. 

As is well known, the solar neutrino deficit can be explained in 
terms of the matter induced resonant Mikheyev-Smirnov-Wolfenstein (MSW)
oscillation \cite{msw} for two choices of masses and mixing 
angles\cite{mix}. Our interest here is in the so called small angle 
solution for which $\Delta m^2_{e\mu} \simeq (0.3-1.0) \times 
10^{-5}$ $eV^2$ and $2 \times 10^{-3} \le \sin^2 2 \theta_{e \mu} \le 2 
\times 10^{-2}$; The atmospheric neutrino deficit could be due to either 
$\nu_\mu \leftrightarrow \nu_\tau$ or $\nu_\mu \leftrightarrow 
\nu_e$ oscillation. Preliminary indications from the electron energy 
distribution in SuperKamiokande favors $\nu_\mu \leftrightarrow \nu_\tau$ 
oscillation. Similarly a preliminary fit to all the atmospheric 
neutrino data (sub-GeV, multi-GeV including the zenith angle dependence) 
seems to require $2 \times 10^{-4} \le m^2_{\mu \tau} ({\rm eV^2}) \le 
10^{-2}$ with $\sin^2 2 \theta_{\mu \tau} \simeq 0.6-1.0$\cite{valle}. Note 
the hierarchical pattern of mass differences. The LSND results require
that $0.3~$eV$^2 \leq \Delta m^2_{e\mu} \leq 10$ eV$^2$ with the mixing 
angle in the few percent range. If we accept the above results, it is clear 
that with only three neutrinos, it is not possible to explain the three
results (i.e. solar, atmospheric and LSND) simultaneously. Therefore
within conventional grand unified theories with three generations, one
may hope to understand only two of the above results. Furthermore,
since hierarchical mass patterns for neutrinos is a generic feature of
theories that implement the see-saw mechanism\cite{grsy}, the solar
and the atmospheric neutrino data appear more amenable to theoretical 
understanding in simple models. 

It is the goal of this 
paper to present a simple grand unified scheme (GUT) that leads to the 
supersymmetric 
standard model at low energies and predict $\Delta m^2$ and $\sin^2 2 
\theta$ values in the above range for $\nu_e \leftrightarrow \nu_\mu$ and 
$\nu_\mu \leftrightarrow \nu_\tau$ sectors so that we have a theoretical
understanding of the solar and the atmospheric neutrino data. We believe 
this result to be significant since we do not use any extra fermions
nor any extra symmetries for the purpose.

The simplest GUT theory that leads naturally to small neutrino 
masses via the see-saw mechanism\cite{grsy} is the SO(10) model where the 
local B-L symmetry is broken by the ${\bf 126 + \overline{126}}$ 
representation. It also has another attractive feature that it leads to 
automatic R-parity conservation so that unwanted (and uncontrolled) 
baryon violating interactions of the MSSM are forbidden and one obtains a 
stable LSP which can act as the cold dark matter of the universe. The 
minimal set of Higgs multiplets needed to break all gauge symmetries of 
the theory while keeping supersymmetry unbroken down to the weak scale 
is: ${\bf 45 + 54}$ (denoted by A and S) ${\bf 126 + \overline{126}}$ 
(denoted by $\Delta$ and $\overline{\Delta}$) and a single ${\bf 10}$, 
denoted by H.

It was shown, sometime ago \cite{babu,lee} that in this minimal 
model, all Yukawa couplings and Higgs vevs responsible for fermion masses 
and mixings (a total of twelve parameters in all in the absence of 
CP-violation) are completely determined by the quark and lepton masses 
and the quark CKM angles. As a result the light and heavy Majorana mass 
matrices for the neutrinos are completely determined except for the 
overall scale $v_R$, the scale of B-L symmetry breaking, provided one 
assumes the simple see-saw formula (to be called type-I see-saw 
formula\cite{grsy}) 
\be
M_\nu=-M^D_{\nu}~M^{-1}_{N_R}~[{M^D_{\nu}}]^T. 
\ee
This enables a complete prediction of neutrino mixing angles and any two 
neutrino mass ratios. Every choice for the signs of the various charged 
fermion masses lead to distinct scenarios and separate predictions. It was 
found that there 
were predictions that could accommodate only small angle MSW solution to 
the solar neutrino puzzle but not the atmospheric neutrino puzzle. The 
reason was that the maximum value for $\sin^2 2 \Theta_{\mu \tau}$ mixing 
angle predicted by this model was less than 0.3 or so, where as the present 
99\% confidence level fits seem to require $\sin^2 2 \Theta_{\mu \tau} 
\simeq 0.60$ or higher\cite{valle}.

One may try to take advantage of the fact that in most left-right and 
SO(10) models, a generalized see-saw formula for neutrino masses holds 
\cite{goran} (to be called the type II see-saw formula):
\be
M_\nu=~f v_l -M^D_{\nu}~M^{-1}_{N_R}~[M^D_{\nu}]^T, 
\ee
(where $v_L \simeq \lambda {V^2_{wk} \over v_R}$ and is induced as long as 
there are {\bf 54} dimensional Higgs multiplets in the theory) 
and see if it is possible to obtain larger values for 
$sin^22\theta_{\mu\tau}$. Such models would have two free parameters, 
$v_R$ and $v_L$. In SUSY GUT models, coupling constant unification 
including threshold corrections puts $v_R$ from $10^{13}$ to $10^{15}$ 
GeV range leading to $v_L \simeq 10^{-2} ~ 
{\rm to}~ 1$ eV for $\lambda \simeq 1$. The two terms then give comparable 
contributions and we have two parameters \{$v_L$,$v_R$\} that determine 
the neutrino masses and mixings. We have made an extensive numerical 
analysis of the predictions of this model 
for neutrino masses and were unable to find any reasonable values
of $v_L$ and $v_R$ which can 
accommodate both the small angle MSW solution to the solar 
neutrino problem as well as the $\nu_\mu \leftrightarrow \nu_\tau$ 
oscillation solution to the atmospheric neutrino puzzle. Thus 
neutrino experiments may play the role similar to the role that proton 
decay experiments played in ruling out minimal non-supersymmetric SU(5) 
model.

We therefore are led to consider a slight generalization of the above 
idea and consider an SO(10) model with two {\bf 10}-dim. Higgs multiplets 
instead of one as in the minimal model.
The rest of the field content is the same. The remainder of the paper 
will be 
devoted to studying the neutrino masses in this model.  The low energy 
theory in this model is the MSSM with the Higgs doublets in general being 
linear combinations of the
doublets in the {\bf 10}'s (denoted $H_{1,2}$) and the $\overline{\bf 126}$.
We will assume the following specific form for them.
\begin{eqnarray}
H_u&=&\alpha_1 H_u({\bf 10}_1) +\alpha_2 H_u(\overline{\bf 126})+\alpha_3 
H_u({\bf 126})  \nonumber \\
H_d&=&\beta_1 H_d({\bf 10}_2)+\beta_2 H_d(\overline{\bf 126})+\beta_3 
H_d({\bf 126}) 
\end{eqnarray}
 How the light doublets arise with this specific form is of course related 
to the difficult problem of 
doublet-triplet splitting in SO(10) models which is not addressed here. 

Let us now discuss how the neutrino mixing angles can be extracted from 
this model. The first point is that the most general Yukawa 
superpotential of the model given by, 
\be
W_Y = h_{i,ab}~ \psi_a~ \psi_b~ H_{i} + f_{ab}~ \psi_a~ \psi_b~ 
\overline{\Delta}, \label{super} 
\ee
where $\psi_a~(a=1,3)$ represent the {\bf 16} dimensional spinors 
corresponding to the three family of fermions. Since SO(10) symmetry 
implies that $h_i$ and $f$ are symmetric matrices, (we ignore CP 
violation from Yukawa sector), we can diagonalize any one of them and we 
have fifteen 
free Yukawa coupling parameters in terms of which the fermion masses and 
mixings are expressed as follows: 
\bea
&& M_u  =  h_1~ v_u + f~\kappa_u~~~~~M_d = h_2~v_d + f~\kappa_d \nonumber \\
&& M_{\nu^D}  =  h~v_u - 3 f~\kappa_u~~~M_l = h_2~v_d-3 f~\kappa_d 
\eea
Using these relations, we find that at the GUT scale, we have
\bea
M_{\nu^D} &= &r_1(M_l-M_d)+ M_u
 \label{lepton} \label{mnud}\\ 
M_{\nu_L\nu_L}&=&r_2 (M_d-M_l) \label{mnul} \\
M_{N_R} &=& r_3 (M_d-M_l)
 \label{mnum}
\eea
where $r_1=\frac{\kappa_u}{4\kappa_d}; r_2= \frac{v_L}{4\kappa_d}$
and $r_3=\frac{v_R}{4\kappa_d}$ and $M_{\nu_L\nu_L}$ is assumed to
denote the $fv_L$ contribution to the neutrino mass matrix. From the above
equations it is clear that we need to supply six parameters to determine
the neutrino masses and mixings and they are the three miximg angles
in the charged lepton mass matrix and $r_i$ ($i=1,2,3$). We demand that
the three charged lepton mixing angles are zero. We then scan the 
parameter space for $r_i$ to see if any desirable solution exists. 

To proceed with this program, first note 
that the above relations between fermion masses hold at the GUT 
scale. So, we extrapolate the observed values of quark and lepton masses 
to the GUT scale, using simple analytic formulae given by Naculich 
\cite{nacu}. We work in a basis where $M_u$ is diagonal and $M_d= V_{ckm} 
D_d V^\dagger_{ckm}$. At the GUT scale the diagonalized values for 
the masses in GeV and the values of the angles are\\ 

\noindent $
m_u=0.0011~~~~
m_c=-0.3785~~~~
m_t=-112.34 \\
m_d=0.00131~~~
m_s=0.0148~~~~
m_b=-1.177 \\
m_e=0.0003~~~~
m_\mu=-0.0699~~~~
m_\tau=1.183~\\
s_{12}=-0.2210~~~
s_{13}= 0.0040~~~
s_{23}= 0.0310
$

\noindent where $s_{12}$ is the Cabbibo angle, $s_{13}$ and $s_{23}$ are 
roughly the $V_{ub}$ and $V_{cb}$ elements of $V_{ckm}$. 

In the basis we are working, ${M}_l$ is diagonal. Furthermore, since the 
signs of the fermion masses are arbitrary, we choose a basis where the 
various fermion masses have the signs as given above. We then use 
Eq.(\ref{mnud}) and Eq.(\ref{mnum}) for 
each of the cases, to obtain the neutrino masses and mixing angles.

Note that we still need to know $v_L\equiv 4r_2\kappa_d$ and $v_R\equiv 
4r_3\kappa_d$. One can use theoretical arguments for the orders of magnitude
of the parameters $v_L$ and $v_R$ that are plausible. Note, for instance that
since the value of the 
induced vev $v_L \simeq {\kappa^2 \over v_R}$, for $v_R$ in the range of 
$10^{13}-10^{16}$ GeV, $v_L \simeq 1~- 10^{-2}$ eV is quite reasonable.  
One way to 
determine $v_R$ is to use the unification constraint as it applies to the 
minimal model. We assume that the theory below the $v_R$ scale is the 
minimal supersymmetric standard model (MSSM). 
Since different choices of the particle spectrum above the intermediate 
scale give different values of $v_R$,  we use another method to 
constrain this parameter.

{\noindent \underbar{Baryogenesis Constraints on the Scale $v_R$}}

A very simple mechanism for baryogenesis in SO(10) models is to generate 
a lepton asymmetry at a high temperature via the decays of the 
right-handed Majorana neutrinos and have this lepton asymmetry converted 
to a baryon asymmetry\cite{fuya} by the sphaleron processes. 
An important necessary condition for this to happen is that at-least one 
of the right handed Majorana neutrinos must have a decay rate that is 
slower than the expansion rate of the universe when $T \simeq M_N$. 
The general formulae are: 
\be
\Gamma_{N_a} \simeq {\sum_{b} h^2_{i,ab} + \alpha f^2_{ab} \over 8 
\pi}~M_{N_a}  \le 1.73~(g^*)^{1/2}~{M^2_{N_a} \over M_{Pl}}  
\label{baryo}. \ee
Since in our model the Yukawa couplings are all predicted in terms of
vev's $v_u, v_d$ and $\kappa_d$, we can obtain 
a lower bound on $M_{N_i}$ if we know the vev's and using the predicted 
value for $f$ matrix, we can then deduce $v_R$. Since in our analysis is 
independent of $v_{u}$ and $v_d$, the only constraint on them is that 
$\sqrt{v^2_u+v^2_d+\kappa^2_u +\kappa^2_d}= 246$ GeV. Using the fact that
we have chosen $r_1=40.3$ (see below), we can get $\kappa_d\simeq 1 $ GeV. 
Using them, we find that $v_R \ge 10^{14}$ GeV. 

{\noindent \underbar{Prediction for Neutrino masses and mixings}}

Using Mathematica we have scanned over all possible choices for the signs 
of the charged fermions to see if there is a prediction that fits the 
requirements of both the Solar and the atmospheric neutrino puzzles. 
In Figures (1) and (2) we 
we plot the $\Delta~m^2_{e\mu}$ and $\sin^2 2 \Theta_{e\mu}$ as functions 
of $v_L$ and $v_R$. In Figure (3) 
and (4) we plot $\Delta~m^2_{\mu\tau}$ and $\sin^2 2 \Theta_{\mu\tau}$ as 
functions of $v_L$ and $v_R$. Cases \{A,B,C\} have 
$r_3=\{1.81,1.89,1.98\} \times 10^{13}$ respectively. 
We see that $\sin^2 2\Theta_{\mu\tau}$ is the most sensitive function of 
$v_L$ and we find acceptable solutions, displayed in Figure (4):~
$\{m_{\nu_e},~m_{\nu_\mu},~m_{\nu_\tau}\}=-\{0.063,3.087,10.88\}~~
10^{-3}$ in eV and 
\be
U_\nu=\pmatrix{-0.989 & -0.081 & -0.123 \cr
-0.147 & 0.539 & 0.829 \cr
0.001 & -0.838 & 0.545}
\ee
Which gives $\sin^2 2 \Theta_{e\mu}=2.8\times 10^{-2}$ and $\sin^2 2 
\Theta_{\mu \tau}=0.84$ and the Yukawa couplings $h_{1/2}$ and $f$ are, 
\be
h_1={1 \over v_u}\pmatrix{-0.064 & 0.123 & 0.181 \cr
0.123 & -3.721 & 1.488 \cr
0.181 & 1.488 & -17.285}
\ee
\be
h_2={1 \over 100~v_d} \pmatrix{0.155 & -0.228 & -0.338 \cr
-0.228 & -0.775 & -2.769 \cr
-0.338 & -2.769 & ~~-58.598}\ee
\be
f={1 \over 1000~k_d}\pmatrix{0.405 & -0.761 & -1.126 \cr
-0.761 & 20.735 & -9.230 \cr
-1.126 & -9.230 & -589.692}
\ee
Where we had $r_1=40.3$, $r_2=3.15 \times 10^{-12}$, $r_3=1.89 \times 
10^{13}$. Using this, explicit determination of $h_{1/2}$ in combination 
with 
baryogenesis constraint Eqn.(\ref{baryo}), we obtain the lower limit on 
$v_R \ge 10^{14}$ GeV as stated earlier. A few comments are in order on 
other aspects of the SUSY SO(10) model characterized by the 
superpotential in Eq.(\ref{super})

\noindent(i) The doublet-triplet splitting in this model has the 
non-trivial property that it leads to realistic fermion mass spectrum in 
contrast with the Dimopoulos-Wilczek (DW) mechanism. The point is that in 
the DW case the MSSM doublets arise from {\bf 10} dimensional SO(10) 
multiplets thereby leading to incorrect mass relations ${m_e \over 
m_\mu}={m_d \over m_s}$, which is off by a factor 10 or so. In contrast, 
in our model the low energy MSSM doublets are admixtures of doublets in 
${\bf 10}$ and ${\bf \overline{126}}$ and is therefore free of such 
difficulties.


\noindent(ii) It is also worth emphasizing that, the near maximal mixing 
angle for $\nu_\mu \leftrightarrow \nu_\tau$ sectors needed to explain 
the atmospheric neutrino data is very hard to obtain with the type I 
see-saw formula as has been clear in many studies 
\cite{babu,lee,aichman}. One generally needs heavier vectorlike quarks 
\cite{barr} for this purpose. Thus our analysis would speak in favour of the 
type II see-saw formula which puts constraints on the SO(10) model building.

\noindent(iii) Strictly speaking the prediction of neutrino masses is  
sensitive to the renormalization of the see-saw formula\cite{bludman}. 
However, in our case, the Yukawa couplings are so small that (as can be
seen from Eqs. 11, 12 and 13 ) this extrapolation does not noticeably alter
the above predictions at low energies. 

In conclusion, we have shown that the use of a type II see-saw formula in a 
next to minimal SUSY SO(10) model without extra matter multiplets or 
extra symmetries can explain both the solar and atmospheric neutrino 
deficits but not the LSND results. Thus if the LSND results are confirmed by 
KARMEN experiment, this class of SUSY SO(10) models (minimal and next to 
minimal) cannot accomodate it simultaneously with the solar and the 
atmospheric neutrino results and alternative theoretical frameworks must
be investigated.

\noindent This work was supported by the National Science Foundation 
grant number. PHY-9421386. We are grateful to Bhaskar Dutta for critical 
comments on an earlier version of the paper.

\noindent FIGURE CAPTIONS:\\
FIG.1: Predicted $\Delta m^2_{e\mu}$ for various $v_L$ and $v_R$.\\
FIG.2: Predicted $\sin^2 2\Theta_{e\mu}$ for various $v_L$ and $v_R$.\\
FIG.3: Predicted $\Delta m^2_{\mu\tau}$ for various $v_L$ and $v_R$.\\
FIG.4: Predicted $\sin^2 2\Theta_{\mu\tau}$ for various $v_L$ and $v_R$.\\
\begin{figure}[h]
\begin{center}
\epsfxsize= 6cm
\epsfysize= 4.5cm
\mbox{\hskip -1.0in}\epsfbox{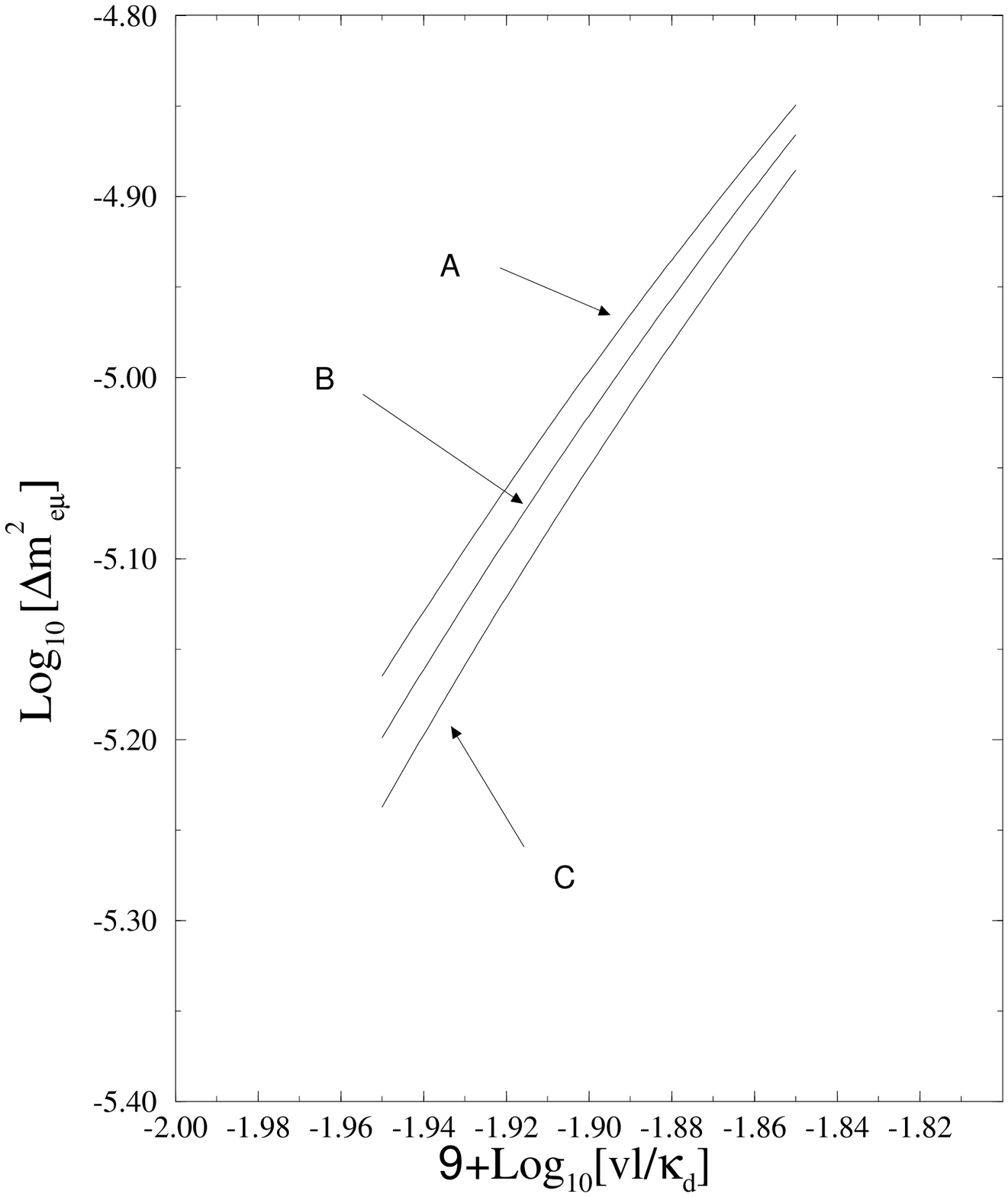}
\caption{} \end{center} 
\end{figure}

\begin{center}
\begin{figure}[h]
\begin{center}
\epsfxsize=6cm
\epsfysize=4.5cm
\mbox{\hskip -1.0in}\epsfbox{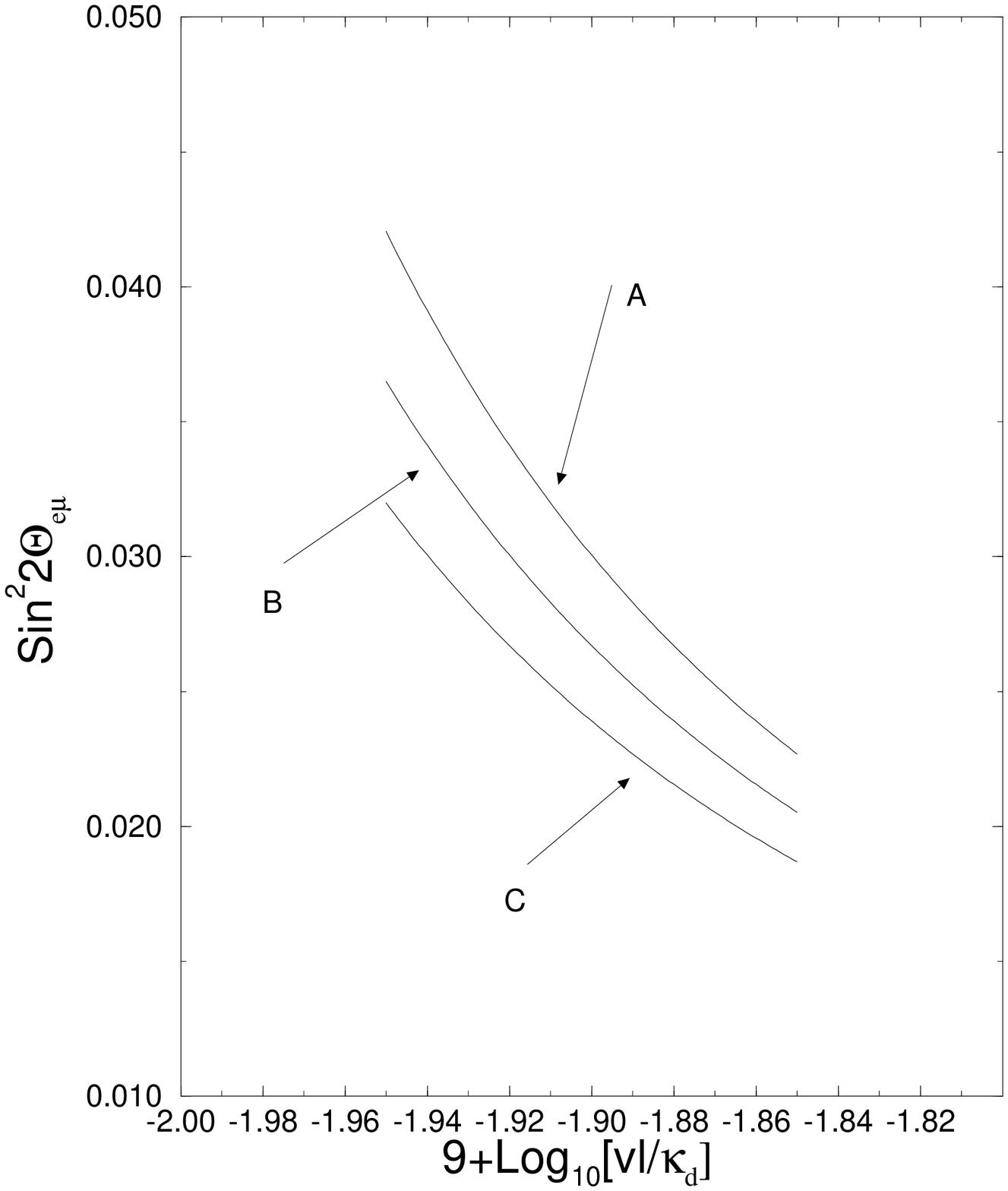}
\caption{} \end{center}
\end{figure}
\end{center}

\begin{figure}[h]
\begin{center}
\epsfxsize=6cm
\epsfysize=4.5cm
\mbox{\hskip -1.0in}\epsfbox{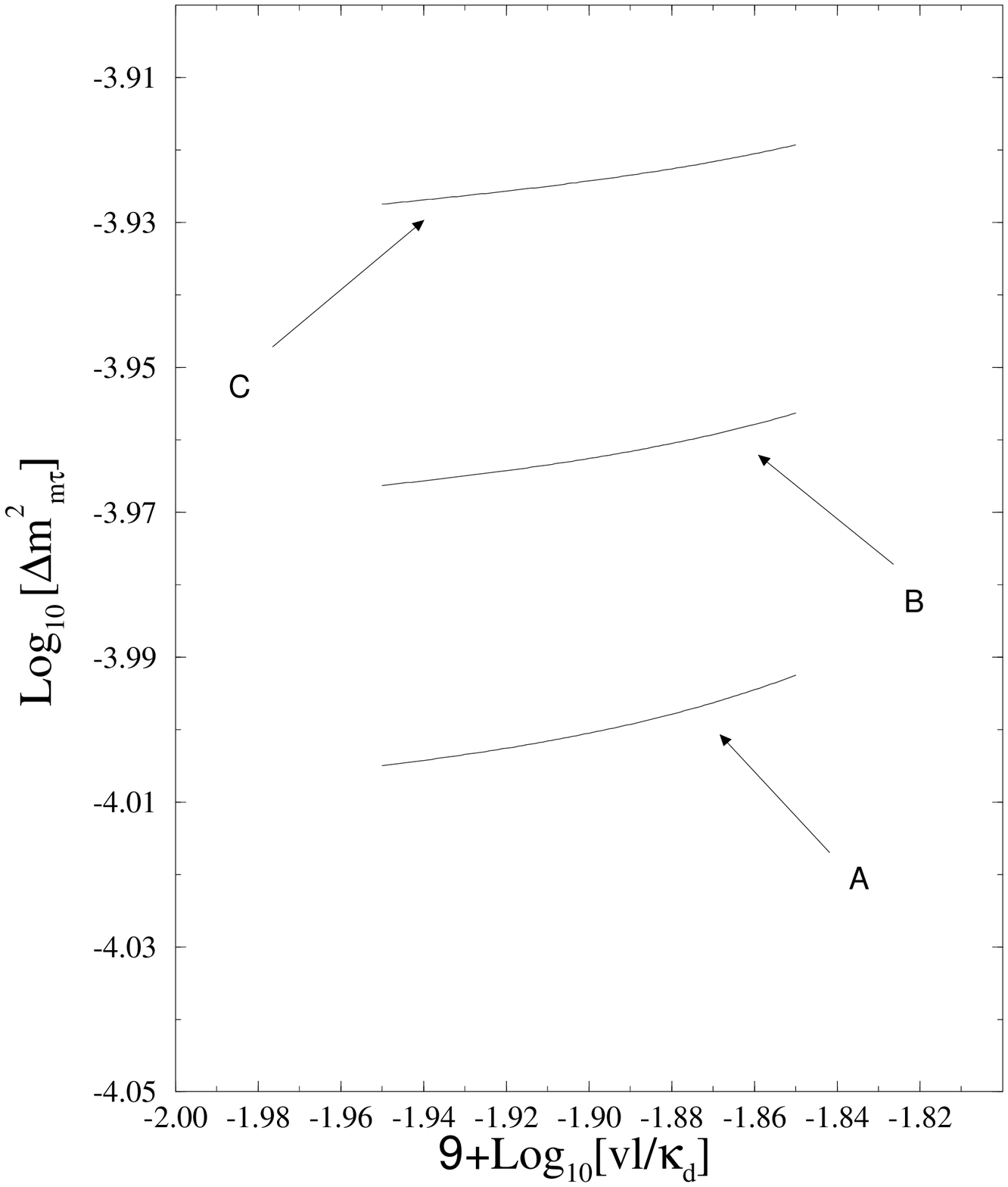}
\caption{} \end{center}
\end{figure}

\begin{figure}[h]
\begin{center}
\epsfxsize=6cm
\epsfysize=4.5cm
\mbox{\hskip -1.0in}\epsfbox{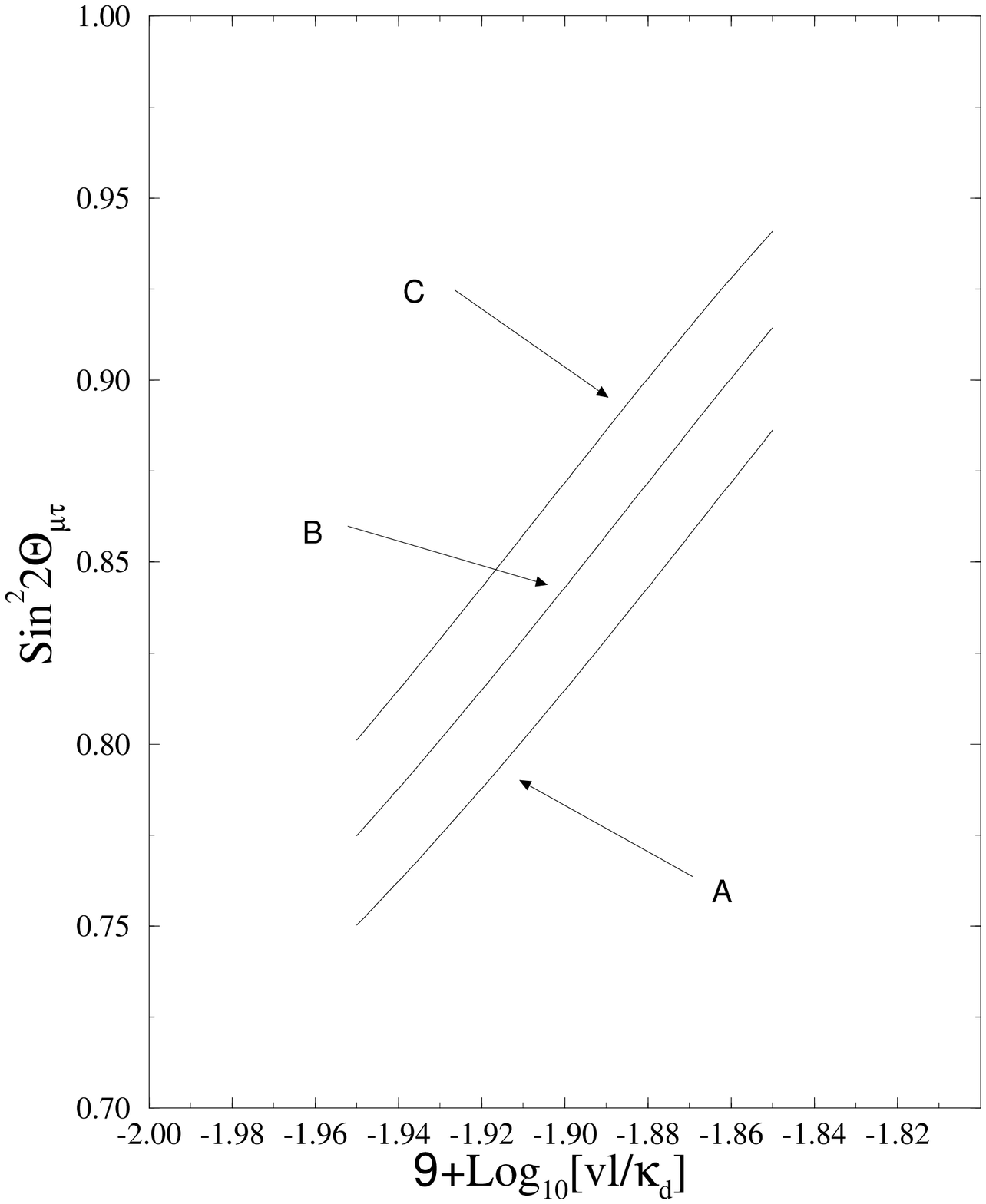}
\caption{} \end{center}
\end{figure}

\end{document}